\documentclass[%
 reprint,
superscriptaddress,
 amsmath,amssymb,
 aps,
]{revtex4-1}

	\usepackage{bm}
	\usepackage{braket}
	\usepackage{dsfont}

 	\usepackage{graphicx}
	\usepackage[caption=false]{subfig}
	\usepackage[export]{adjustbox}

	\newcommand{\vect}[1]{\boldsymbol{#1}}		
	\newcommand{\op}[1]{\hat{\boldsymbol{#1}}}	

			\graphicspath{ {FinalDiagrams/} }

\begin{document}

\title{Valley-polarised tunnelling currents in bilayer graphene tunnelling transistors}
\author{J. J. P. Thompson}
\email{J.J.P.Thompson@bath.ac.uk}
\affiliation{Department of Physics, University of Bath, Claverton Down, Bath BA2 7AY, United Kingdom}
\author{D. J. Leech}
\affiliation{Department of Physics, University of Bath, Claverton Down, Bath BA2 7AY, United Kingdom}
\affiliation{Centre for Fine Print Research, University of the West of England, Bristol BS3 2JT, United Kingdom}
\author{M.~Mucha-Kruczy\'{n}ski}
\affiliation{Department of Physics, University of Bath, Claverton Down, Bath BA2 7AY, United Kingdom}
\affiliation{Centre for Nanoscience and Nanotechnology, University of Bath, Claverton Down, Bath BA2 7AY, United Kingdom}

\begin{abstract}
We study theoretically the electron current across a monolayer graphene/hexagonal boron nitride/bilayer graphene tunnelling junction in an external magnetic field perpendicular to the layers. We show that change in effective tunnelling barrier width for electrons on different graphene layers of bilayer graphene, coupled with the fact that its Landau level wave functions are not equally distributed amongst the layers with a distribution that is reversed between the two valleys, lead to valley polarisation of the tunnelling current. We estimate that valley polarisation $\sim 70\%$ can be achieved in high quality devices at $B=1$ T.  Moreover, we demonstrate that strong valley polarisation can be obtained both in the limit of strong-momentum conserving tunnelling  and in lower quality devices where this constraint is lifted.
\end{abstract}
\maketitle

\section{Introduction}

Electron tunnelling through a potential barrier is one of the most widely known physical consequences of quantum mechanics, responsible for effects as varied as nuclear fusion in stars, radioactive decay or spontaneous DNA mutation \cite{Trixler}. In particular, the probability of successful tunnelling decays exponentially with the width of the barrier, an effect best visualized in scanning transmission microscopy where moving the conducting tip away from the sample leads to rapidly decaying tunnelling currents, hence allowing for imaging of the corrugation of the sample surface \cite{STM}. Recently, the limit of single-atomic-layer barrier thickness has been achieved in van der Waals (vdW) heterostructures using, first, atomically thin graphene and hexagonal boron nitride (hBN) \cite{britnell_nl_2012,britnell_sci_2012}, as the electrode and barrier respectively, and later other two-dimensional atomic crystals \cite{georgiou_natnano_2013}. The resulting tunnelling transistors offer a solution to the graphene `band gap problem' --- the lack of a band gap in the conical electronic dispersion of the material \cite{britnell_sci_2012}. Moreover, in devices with ultra-high quality interfaces, momentum-conserving tunnelling was demonstrated, leading to negative differential resistance \cite{britnell_natcom_2015, mishchenko_natnano_2014, greenaway_natphys_2015, lane_apl_2015, fallahazad_nanolett_2015, kang_ieee_2015} and valley polarization due to an in-plane magnetic field \cite{wallbank_sci_2016}. It was also shown that electron tunnelling in vdW heterostructures can be accompanied by excitation of various quasiparticles, for example, phonons \cite{phononassist} or magnons \cite{magnons} and influenced by defects in the tunnel barrier \cite{phonondefect, defect}. Finally, moir\'{e} superlattice effects can be used to engineer the electronic densities of states of the electrodes \cite{li_natphys_2010, yankowitz_natphys_2012, leech_physrevapplied_2018, suggestedpaper}.

Here, we study theoretically the tunnelling current flowing between bilayer graphene (BLG) and monolayer graphene (MLG) electrodes through a hBN barrier, in the presence of a magnetic field perpendicular to the atomic layers. The impact of the applied magnetic field is two-fold: firstly, the electronic density of states is modified due to Landau quantisation and second, layer polarisation of the low-energy Landau levels in BLG \cite{mccann_prl_2006} leads to  efficient generation of valley polarisation \cite{Rycerz, valleyreview}. We show here that valley polarisation of order unity is possible, in magnetic fields as low as $\sim 1$ T, and that choice of the valley quantum number of the tunnelling current can be made electronically without reversing or changing the magnitude of the magnetic field. While the largest valley polarisation can be achieved in high quality devices in which tunnelling electrons conserve both energy and momentum, our results suggest that even in the absence of momentum conservation, polarisation $\sim 70\%$ can be achieved at $B = 1$ T.

\section{Device Description And Tunnelling Matrix Element}

\begin{figure} 
\centering
\includegraphics[clip,width=0.95\columnwidth]{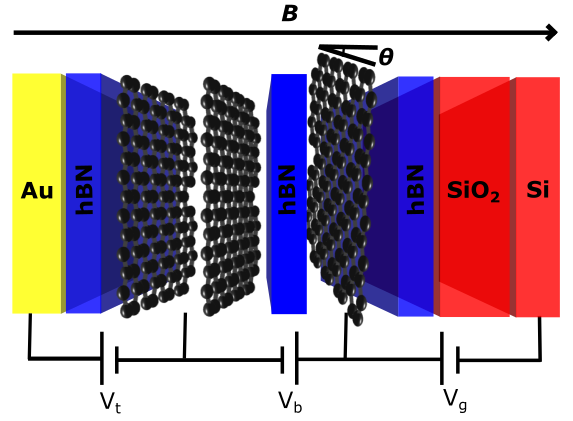}\\
\caption{(Color online) Schematic of the tunnelling device discussed here. Also shown are the tuning potentials $V_b$, $V_g$ and $V_t$ and direction of the applied magnetic field $B$.}
\label{fig:figure1}
\end{figure}

The schematic of the device we study is shown in Fig. \ref{fig:figure1} with the assumed direction of the magnetic field, $B$. We consider BLG, two layers of carbon atoms arranged in regular hexagons and stacked in a Bernal (\textit{AB}) formation, adjacent to a few-layer hBN sheet with a monolayer graphene (MLG) sheet placed onto the opposite face, such that the hBN acts as a tunnelling barrier between the two graphene materials. The BLG and MLG electrodes can be rotationally misaligned  so that their crystallographic directions are rotated by an angle $\theta$ \cite{kim_nanolett_2016}.  Following experimental device architectures \cite{ dean_natnano_2010, mishchenko_natnano_2014, wallbank_sci_2016}, we assume that both MLG and BLG are encapsulated with hBN. We additionally assume that the misalignment between graphene electrodes and hBN (both the barrier as well as the top and bottom encapsulating layers) is large so that moir\'{e} effects like miniband formation \cite{yankowitz_natphys_2012} or lattice relaxation \cite{refpaper}, important for highly-aligned interfaces \cite{Marcinpaper17}, can be neglected.  Finally, a silicon back gate which enables doping of the tunnelling electrodes is located on the side of MLG, separated by  SiO$_2$, while a gold top gate is attached adjacent to the BLG electrode. 

In order to find the tunnelling current through the device we use Bardeen's formalism \cite{britnell_sci_2012, greenaway_natphys_2015, Bardeen, simmons, li, Feenestra}, which utilises the wave functions of the source and drain electrodes to model the tunnelling probability. The matrix element, $M(\varepsilon)$, associated with the probability of an electron with energy $\varepsilon$ tunnelling through the barrier (which we take to lie in the \textit{xy}-plane), is calculated (up to some constant prefactor with dimension of energy $\times$ distance) as
\begin{align} \label{Bardeen}
  M\!\!\propto\!\!&\int\left[\Psi_\text{S}(\vect{r},\varepsilon)\dfrac{\partial \Psi_\text{D}^*(\vect{r},\varepsilon)}{\partial z}\!-\! \Psi_\text{D}(\vect{r},\varepsilon)\dfrac{\partial\Psi_\text{S}^*(\vect{r},\varepsilon)}{\partial z}\right]dV ,
\end{align}
where $\Psi_{\text{S}} $ ($\Psi_{\text{D}}$) describes the wave function on the source (drain) electrode and the integration is over the volume of the tunnelling junction. 
In a multilayer electrode, such as bilayer graphene, the overall wave function can be described as a linear combination of the wave functions on the constituent layers. Therefore, in the case of an $N$-layer source electrode we can write
\begin{align}
    \Psi_\text{S}(\vect{r},\varepsilon) = \sum_{i=1}^N n_i \psi_{\text{S}, i}(\vect{r}, \varepsilon),
\end{align}
where $\psi_{\text{S}, i}$ is the electronic wave function of the $i$-th layer of the source electrode and $|n_i|^2$ describes the relative occupation of the $i$th layer by an electron in state $\Psi_\text{S}$.

Assuming a clean sample, all $\psi_{\text{S}, i}(\vect{r}, \varepsilon)$ are separable into the in-plane and perpendicular components, $\psi_{\text{S}, i}(\vect{r}, \varepsilon) = \varphi_{\text{S}, i} (x,y,\varepsilon) \phi_{S, i}(z, \varepsilon)$. This enables us to decompose the matrix element, Eq. (\ref{Bardeen}), into transverse, $z$, and in-plane, $x, y$, components. We model the transverse component of the wave functions, $\phi_{\text{S},i}(z, \varepsilon)$, as  exponentially decaying, taking care to change the decay constant in the regions corresponding to different materials. In our case, there are two materials to consider: hBN, comprising the tunnelling barrier, and graphene. We assign to them decay constants $c(\varepsilon)$ and $c'(\varepsilon)$, respectively (the decay constants vary as a function of the tunnelling state energy  $\varepsilon$ \cite{simmons}). With these assumptions, we integrate the expression in Eq. (\ref{Bardeen}) in the direction transverse to the barrier. For our BLG source electrode which consists of two graphene layers, the electrons from the layer closest to the barrier tunnel directly to the drain, passing only through one material, hBN. However, the electrons in the layer further from the barrier have to travel an increased distance. Since there are no available states on the other graphene layer the electrons have to pass through, it can be effectively treated as an insulator and so the only mechanism for transport is tunnelling. Integrating over the total width of the barrier for these electrons, defined as the sum of the hBN barrier width, $d$ and the interlayer separation between graphene layers, $d_0$, we obtain an expression depending only on the in-plane components of the wave function and energy,

\begin{align} \label{MatEle}
    M(n_1, n_2, \varepsilon)\propto & \;\; e^{-{c(\varepsilon)}{d}} \bigg[ \int  \varphi^*_\text{D}(x,y,\varepsilon) \\
 \cdot & \begin{pmatrix}
   n_1 & n_2 e^{-{c'(\varepsilon)}{d_0}} 
    \end{pmatrix} \cdot \begin{pmatrix}
    \varphi_\text{S,1}(x,y,\varepsilon)\\ \notag
    \varphi_\text{S,2}(x,y,\varepsilon)
    \end{pmatrix} dA\bigg],
\end{align}
where we have absorbed all normalisation factors into $ \varphi_{\text{S},i}(x, y, \varepsilon)$ and used the fact that in our device the drain electrode is built of only one layer and source of two. Note that the expression in Eq. (\ref{MatEle}) conserves the in-plane electron momentum in the tunnelling process, as is the case in experiments performed on the highest quality devices \cite{britnell_natcom_2015, mishchenko_natnano_2014,  greenaway_natphys_2015, wallbank_sci_2016}. 

While the argument above can be extended to any number of layers in both the source and drain electrodes, the exponential dependence of tunnelling probability on the barrier width means that only tunnelling from/into the first few layers next to the barrier is measurable. For hBN, experimental works \cite{britnell_nl_2012, britnell_sci_2012, greenaway_natphys_2015, wallbank_sci_2016, ROY, Lee, PhysRevB.95.045303} suggest a value of the decay constant $c(0)\approx 5 \text{nm}^{-1}$. In the case of graphene, studies of its role as a barrier in magnetic tunnel junctions  \cite{Friedman, Friedman2, Erve} and between metal contacts \cite{Kuroda, Zhong} showed that it behaves as a strong out-of-plane insulator. In fact, in 
experiments conducted in the absence of a magnetic field and in the presence of a field parallel to the graphene layers, the measured tunnelling current has been well described by assuming that all tunnelling from the further BLG layer is suppressed \cite{lane_apl_2015, wallbank_sci_2016}. For this reason, here, we take the limit $c'= c$, corresponding to the decay through graphene being significant and similar to that through hexagonal boron nitride. However, our conclusions hold for notably smaller $c'$ (we discuss what happens for differing estimates of $c'$ in Appendix A). 
 
\section{Wave functions of graphene electrodes}

In order to obtain the wave functions of electrons in our BLG and MLG electrodes, we use the low-energy description for electrons in these materials, applicable in the vicinity of the inequivalent Brillouin zone corners (valleys) $\vect{K_{\xi}} = \xi (4 \pi/3 a,0)$, where $\xi = \pm 1$ and the graphene lattice constant $a = 2.46$ \AA. 
A single graphene layer consists of two sublattices, $A$ and $B$, and in the case of BLG, an effective low-energy model can be constructed using Bloch states, $\phi(A1)$ and $\phi(B2)$, formed from $p_z$-orbitals on the non-dimer sites (sites which do not have a neighbour directly above/below them)  \cite{mccann_prl_2006} which we refer to as $A1$ and $B2$ with the labels 1 and 2 corresponding to the layer closer and further from the barrier, respectively. For an electron with momentum $\vect{p} = (p_x, p_y)$ measured from the centre of valley $\boldsymbol{K_\xi}$, the resulting Hamiltonian, written in the basis $\left\{\phi(A1),\, \phi(B2)\right\}^T$ in $\boldsymbol{K_+}$ and $\left\{\phi(B2),\, \phi(A1)\right\}^T$ in $\boldsymbol{K_-}$ , is
\begin{align} \label{BLGham}
\op{H}_{\text{BLG}} &= -\dfrac{v^2}{\gamma_1}\! \begin{pmatrix}
0 &  {\pi^\dagger} ^2 \\
  \pi^2 & 0
\end{pmatrix} + \op{H}_u, \\ \notag
 \op{H}_u &= \frac{\xi u}{2}\begin{pmatrix}
1& 0 \\
 0 &-1 
\end{pmatrix}\! +\! \frac{\xi u v^2}{\gamma_1^2}\!\begin{pmatrix}
 \pi^\dagger \pi & 0 \\
 0 & \pi \pi^\dagger 
 \end{pmatrix}, \\ \notag
 \pi &= p_{x} + i p_{y} .
\end{align}
 Above, the velocity, $v \approx 10^{6}$ ms$^{-1}$, is related to the in-plane nearest-neighbour hopping while $\gamma_{1} \approx 0.38$ eV is the vertical interlayer coupling. The term $\op{H}_u$ captures the effect of energy difference between sites on different layers, $u$ (interlayer asymmetry), due to the electric field perpendicular to the BLG electrode induced by the applied voltages.

Misalignment between the MLG and BLG electrodes, generated by a small clockwise rotation of the MLG sheet about the $z$-axis by angle $\theta$ , leads to an identical rotation between the corresponding Brillouin zones. As a result of this rotation, the position of MLG valley centres is offset from that of BLG by a vector  $\vect{\Delta K_\xi} = \left( \Delta K_{\xi}^{x}, \Delta K_{\xi}^{y} \right) = (\vect{\hat{1}}-\vect{\hat{R}_\theta}) \vect{K_\xi}$, where $\vect{\hat{R}_\theta}$ is the anti-clockwise rotation operator. Taking into account this shift as well as the rotation between the two materials, electrons in the MLG electrode are described by a Hamiltonian
\begin{align} \label{MLGHam}
    \op{H}_{\text{MLG}} &= v \begin{pmatrix} 0 & (\pi^{\dagger} + \widetilde{\pi}^{\dagger})e^{-i \theta} \\ (\pi +\widetilde{\pi})e^{i\theta}& 0 \end{pmatrix},
\end{align} 
which acts on the basis  $\left\{\phi(A),\, \phi(B)\right\}^T$ for the $\vect{K}_+$ valley and $\left\{\phi(B),\, -\phi(A)\right\}^T$ in the $\vect{K}_-$ valley and $\widetilde{\pi}=\hbar (\Delta K_{\xi}^{x} +i \Delta K_{\xi}^{y})$. 

We include the magnetic field $\boldsymbol{B}$ applied perpendicular to the graphene planes in Eq. (\ref{BLGham}) and (\ref{MLGHam}) by using the Peierls substitution, $\vect{p} \rightarrow \vect{p} + e\vect{A}$, and the Landau gauge $\vect{A} = (0, -Bx, 0)$. As a result, the operators $\pi$ and $\pi^\dagger$ become lowering and raising operators, respectively, for functions built of quantum harmonic oscillator states along the $x$-direction, $\phi_m(x)$, and plane waves along the $y$-direction,
\begin{align} \notag
 &\pi \phi_m(x) e^{i k_y y} = -i\dfrac{2\hbar}{\lambda_B}\sqrt{m}  \phi_{m-1}(x)e^{i k_y y}, \\ \notag
  &\pi^\dagger  \phi_m(x) e^{i k_y y} = i\dfrac{2\hbar}{\lambda_B}\sqrt{m+1} \phi_{m+1}(x)e^{i k_y y}, \\ \notag
  &\phi_{m}(x) = \! A_{m} \exp\left[ - \frac{1}{2 \lambda_{B}^{2}} (x \! - \! X)^{2} \right]\mathcal{H}_{m} \left( \frac{1}{\lambda_{B}} (x - X) \right) , \\  \label{raising}
  & A_{m} = \frac{1}{\sqrt{2^{m}m!\sqrt{\pi}\lambda_{B}L}}, 
\end{align} 
where $\lambda_{B} = \sqrt{\hbar/eB}$ is the magnetic length, $\mathcal{H}_{m}(x)$ is a Hermite polynomial of order $m$, $L$ is the dimension of the flake along the $y$-direction and  $X = \lambda_{B}^2 k_y$ is the position of the centre of cyclotron orbit of an electron with wave vector $k_y$.
 Using Eq. (\ref{raising}), the energies of the BLG Landau levels, can be expressed as \cite{mccann_prl_2006}
 \begin{align}\notag
  &\varepsilon_{0,\xi}=  \dfrac{\xi u}{2}, \\ 
      &\varepsilon_{1,\xi}=\dfrac{\xi u}{2} - \xi\eta^2 u,  \\
      &\varepsilon_{m,s,\xi}=-\frac{\eta^2 \xi u}{2} +s \sqrt{ (\varepsilon^0_{m})^2 +\frac{1}{4}u^2}, & m \geq 2, \nonumber
\end{align}
where $\eta = \sqrt{2}v \hbar/\lambda_{B} \gamma_{1}$,  $\varepsilon_{m}^0 = \gamma_1 \eta^2  \sqrt{m(m-1 )}$ and $s=\pm 1$ is the conduction/valence band index in BLG, defined when $m \geq 2$.
 The corresponding wave functions are
\begin{align} \label{waveblg}  &\psi_{0} =
  \begin{pmatrix} \notag
  \phi_{0} \\  
0
\end{pmatrix} e^{i k_{y}  y},  \\ \notag
&\psi_{1} =
  \begin{pmatrix}
  \phi_{1} \\  
0
\end{pmatrix} e^{i k_{y}  y},  \\ \notag
&\psi_{m,s}^{\xi} =\dfrac{1}{\sqrt{2 C}} \begin{pmatrix}
  C_1 \phi_{m} \\  C_2\phi_{m-2}
\end{pmatrix} e^{i k_{y}  y},  & m \geq 2,
\\ 
&C_1 = \varepsilon_{m}^0 , \quad C_2 = \left[\varepsilon_{m,s,\xi}- \xi \frac{u}{2} +\xi u m \eta^2\right],
\end{align} 
  where $C= C_1^2 +C_2^2$ is the normalisation constant.

In rotated monolayer graphene, the displacement of the momentum-origin modifies the harmonic oscillator state, $\tilde{\phi}_{n}(x)$,
\begin{equation}
\begin{split}
    \tilde{\phi}_{n}(x)  =\, &  A_{n} \exp\left[ - \frac{1}{2 \lambda_{B}^{2}} (x \! - \! \tilde{X})^{2} - i \Delta K_{\xi}^{x} (x - \tilde{X}) \right]  \\ 
& \times\mathcal{H}_{n} \left( \frac{1}{\lambda_{B}} (x - \tilde{X}) \right),
\end{split}
\end{equation}
and shifts the cyclotron orbits to $\tilde{X} = \lambda_{B}^{2}( k_{y}+\Delta K_{\xi}^{y})$. The resulting energy levels in MLG are
\begin{align} \notag
&\varepsilon_{0} = 0,  \\
    &\varepsilon_{n,s'} = s' \left(\sqrt{2} v \hbar / \lambda_{B}\right) \sqrt{n}, &n\geq 1,
\end{align}
where $s'=\pm 1 $ is the conduction/valence band index in MLG, defined when $n\geq 1$.
Therefore, the wave function corresponding to the $n$-th Landau level in rotated MLG can be written as
\begin{align} \notag
&\tilde{\psi}_{0} = \begin{pmatrix} 
\tilde{\phi}_{0} \\ 0 \\
\end{pmatrix}• e^{i k_{y} y}, \\  \label{wavemlg}
&\tilde{\psi}_{n,s'} = \dfrac{1}{\sqrt{2}} \begin{pmatrix}
\tilde{\phi}_{n} \\ - s' i e^{i \theta} \tilde{\phi}_{n - 1} \\
\end{pmatrix}• e^{i k_{y}.y}, & n\geq 1.
\end{align}

In MLG, all the Landau levels have an additional four-fold degeneracy due to spin and valley. Moreover, the $n=0$ Landau level is positioned at the Dirac point and its energy does not depend on the magnetic field. In BLG, for $u=0$, in addition to the valley and spin degeneracies of each level, both the $m=0$ and $m=1$ Landau levels sit at the neutrality point, leading to an unusual eight-fold degenerate zero-energy state. Non-zero $u$ lifts both the $m=0, 1$ and valley degeneracies.

In both MLG and BLG, the wave functions are distributed asymmetrically between the two sublattices concerned ($A$, $B$ in MLG and $A1$, $B2$ in BLG). In particular, the electrons in the $n=0$ MLG level and $m=0,1$ BLG states occupy only one of the sublattices \cite{note1}.
For the case of BLG, this results in two states that in the $\vect{K_+}$ valley are located only on layer 1 and in the $\vect{K_-}$ valley only on layer 2. In the vertical tunnelling transistor geometry like in Fig. 1, we expect electrons from layer 2 to have a smaller chance of tunnelling through the barrier than electrons from layer 1, due to the additional effective barrier thickness. As a result, more electrons from BLG $\vect{K_+}$ valley will tunnel through than from the $\vect{K_-}$ valley, leading to valley-polarized current arriving in the MLG drain electrode.

In order to quantify valley polarisation of the tunnelling current, we use the wave functions from Eq. (\ref{waveblg}) and (\ref{wavemlg}) to compute the tunnelling matrix element, Eq. (\ref{MatEle}) (see Appendix B for more details). The Landau level wave functions of BLG are already written so that their components correspond to different layers and we can identify the states in Eq. (\ref{waveblg}) with $( \varphi_\text{S,1}(x,y,\varepsilon),\; \varphi_\text{S,2}(x,y,\varepsilon))^T$ in Eq. (\ref{MatEle}). However, although the MLG wave functions are also written as spinors in Eq. (\ref{wavemlg}), both of their components correspond to wave function amplitudes on sublattices in the same layer. Hence, for a given Landau level state we take $\varphi_{D}=\chi_A + \chi_B $, where $(\chi_A, \; \chi_B)^T$ is the corresponding spinor in Eq. (\ref{wavemlg}).

In clean samples, with a small misalignment angle between the source and drain electrodes the shift $\vect{\Delta K_\xi}$ is small and for all the dominant processes the valley quantum number is conserved in the tunnelling process \cite{greenaway_natphys_2015,wallbank_sci_2016}. Therefore, we  use Fermi's golden rule to relate the tunnelling matrix element to current of electrons originating in the $\vect{K_\xi}$ valley of BLG,
\begin{align} \label{currentintro}
I_\xi= &  \frac{4 \pi e}{\hbar}\sum_{n,m,s',s} \int_{\mu_\text{BLG}}^{\mu_\text{MLG}+\Delta}  |M_{n,m,\xi}^{s, s'}(\varepsilon)|^{2} \\ \notag
&\times D_{\text{BLG}} (\varepsilon, \varepsilon_{m,s,\xi}) D_{\text{MLG}} (\varepsilon- \Delta, \varepsilon_{n,s}) d\varepsilon,
\end{align}
where we have already taken the spin degeneracy into account.
 We measure the energy $\varepsilon$ from the charge neutrality point of the BLG  electrode while $\mu_\text{BLG}$ and $\mu_\text{MLG}$ represent the distance in energy between the charge neutrality point and the chemical potential in the BLG and MLG electrode, respectively. Finally, we define $\Delta$ as the shift between the source and drain neutrality points  such that, in the low temperature limit, the local chemical potentials in the source and drain electrodes, $\mu_\text{BLG}$ and $\mu_\text{MLG} + \Delta$ respectively, determine the energy window within which tunnelling processes can occur, while the number of initial and final states at a given energy is provided by the densities of states $D_\text{MLG}$ and $D_\text{BLG}$ in the monolayer and bilayer graphene, respectively. 
 In a device with high quality layers, free from defects and in a quantizing external magnetic field, these densities of states consist of a series of sharp peaks at the energies of the Landau levels. We model the latter using a Lorentzian shape with the same full width at half maximum for all Landau levels, 2 meV and 4 meV for $B=1$ T and $B=4$ T, respectively, following previous experimental works \cite{greenaway_natphys_2015,taychat}  and theoretical considerations \cite{funk2015}. Finally, we define the valley polarisation, $P$,  of the tunnelling current,

\begin{equation}
    P = \frac{I_{+} - I_{-}}{I_{+} + I_{-}}.
\end{equation}
Because our tunnelling matrix element, Eq. (1), is defined up to a proportionality constant, values of tunnelling current in this paper are given in arbitrary units. Polarisation, however, as a ratio of currents, does not depend on  that constant itself.

We set the thickness of the hBN barrier (separation between the MLG and BLG) to $d=13$ $\text{\AA}$ and the interlayer distance in BLG as $d_0=3.3$ \AA. We also relate the energies $\varepsilon$, $\mu_\text{BLG}$ and $\mu_\text{MLG}$ to the applied voltages $V_t$, $V_b$ and $V_g$ (see Fig. 1) through the electrostatic equations 
\begin{align}
&V_{b} = \frac{1}{e} \left[ \mu_\text{BLG} - \mu_\text{MLG} - \Delta \right], \nonumber \\ 
 &V_g = -\frac{e(n_\text{MLG}\!+\!n_\text{BLG}+\!n_{\text{Au}})(d_{\text{SiO}_2}\epsilon_{\text{hBN}}\!+\!d_{\text{hBN}}\epsilon_{\text{SiO}_2})}{\epsilon_{\text{hBN}}\epsilon_{\text{SiO}_2}\epsilon_0}, \nonumber\\
   & V_t = \frac{-en_\text{Au}d_{\text{Top}}}{\epsilon_0 \epsilon_\text{hBN}},
\end{align}
discussed in more detail in Appendix C. We define $n_\text{MLG}, n_\text{BLG}$ and $n_\text{Au}$ as the carrier densities on the MLG, BLG and gold electrodes respectively. The distance between the gold top gate and BLG, $d_{\text{Top}}$, is set as 30 nm in our numerical calculations. Furthermore,   $d_\text{hBN}$ and $d_{\text{SiO}_{2}}$ represent the  thicknesses of the hBN and SiO$_2$ substrates, which, following previous experimental works, we set as 30 nm and 300 nm respectively. Finally, $\epsilon_0$ is the permittivity of free space while $\epsilon_\text{hBN} \approx 3 $ and $\epsilon_{\text{SiO}_2} \approx 3.9$ are the relative permittivities of hBN and $\text{SiO}_2$. We also take into account that the electric field between the graphene layers of BLG induces the interlayer asymmetry $u$ which we compute self-consistently,
\begin{align}
    u = -\frac{e^2 d_0 (n_\text{Au} + n_{\text{BLG},2}(u))}{\epsilon_0},
\end{align}
where $n_{\text{BLG},i}(u)$ is the carrier density on the $i$-th layer of BLG. For a given interlayer asymmetry, we compute the electronic wave functions for all Landau levels included in the calculation and their distributions on the atomic sites (while the number of Landau levels considered depended on the magnetic field and applied voltage range, we checked the convergence of our results in all cases). For each Landau level, we use the square of the wave function amplitude on the site $B2$ to obtain the contribution to $n_{\text{BLG},2}$ from that level. We then determine the unique value of $u$ for which Eq. (15) is fulfilled.
\section{Momentum-Conserving tunnelling}
\subsection{Total Tunnelling Current at $B=1$ T}
Our simulation of the total tunnelling current, $I = I_+ + I_-$, between the BLG and MLG electrodes, produced using Eq. (12), is shown in Fig. 2. We indicate the boundaries of regions corresponding to fixed lowest filling factors $\nu_\text{MLG}$ and $\nu_\text{BLG}$ in MLG and BLG respectively with the grey lines and label these regions as $(\nu_\text{MLG}, \nu_\text{BLG})$ \cite{footnotell}. For momentum-conserving tunnelling, the strength of the coupling is dictated by the magnitude of the applied magnetic field, relative orientation of the electrodes and Landau level indices of the involved electronic states.

\begin{figure} [!h]
\subfloat{  \includegraphics[clip,width=0.97\columnwidth]{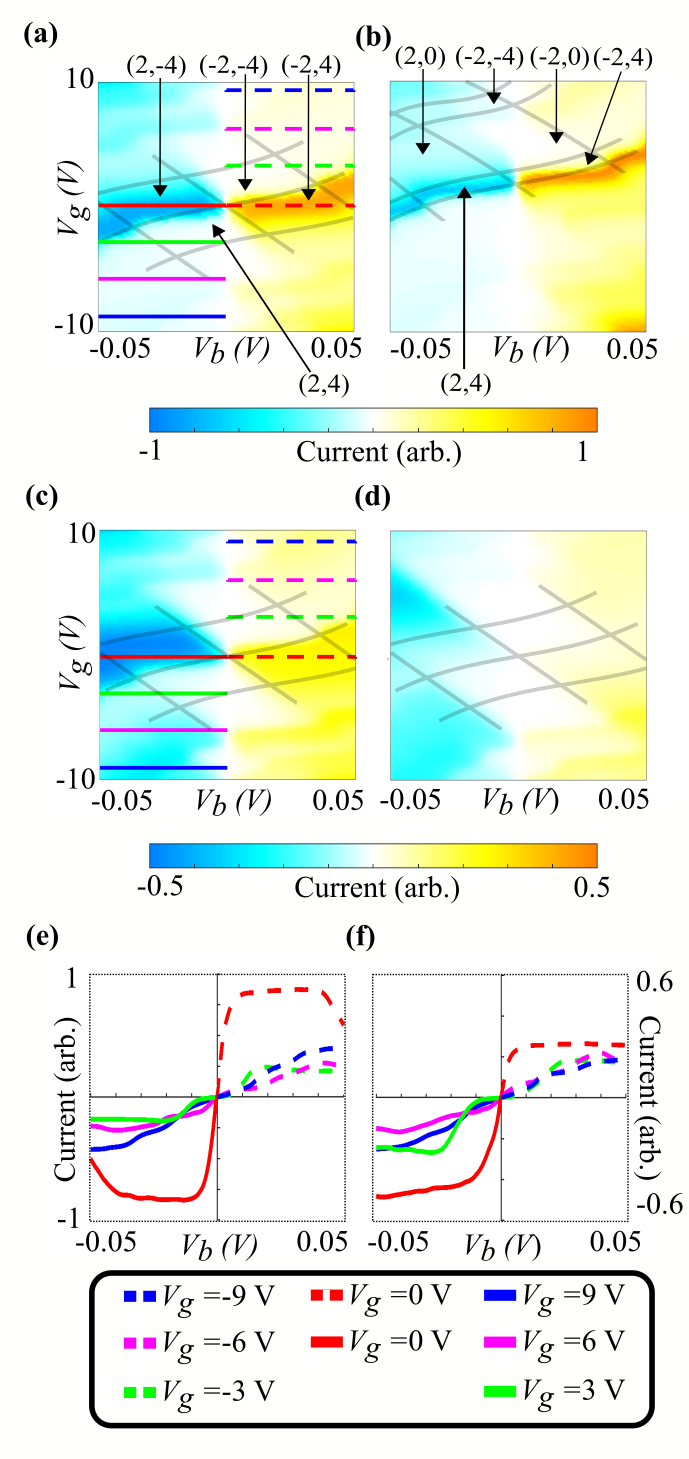} }
\caption{(Color online)  Total tunnelling current at $B = 1$ T  as a function of the bias and gate voltages, $V_b$ and $V_g$, and for (a) $\theta=0^\circ$  and $V_t = 0$ V (b) $\theta=0^\circ$ and $V_t = 0.5$ V, (c) $\theta=0.25^\circ$  and $V_t = 0$ V and (d) $\theta=0.5^\circ$  and $V_t = 0$ V. Grey lines in (a) and (b) indicate the boundaries of regions of constant filling factors labelled as $(\nu_\text{MLG},\nu_\text{BLG})$ with the first (second) filling factor corresponding to monolayer (bilayer) graphene. For clarity, we do not show labels of these regions in panels (c) and (d) for which they are the same as in (a).  Panels (e) and (f) show current curves corresponding to the lines marked in (a) and (c), respectively, with changing $V_b$ and constant $V_g$ from -9 V to 9 V in steps of 3 V. }
\label{fig:figure2}
\end{figure}
In panel (a), we show the current for $V_t=0$ V and ideally aligned electrodes, $\theta=0^{\circ}$. For $V_g = V_b =V_t=0$ V, the chemical potentials in the BLG and MLG electrodes, $\mu_\text{BLG}$ and $\mu_\text{MLG}+\Delta$, are located at their respective neutrality points, which are at the same energy, resulting in zero tunnelling current. As the bias voltage is increased, a shift between the local chemical potentials in the two electrodes is induced. This opens an energy window, within which electrons occupying states in one electrode can tunnel into empty states at the same energy in the other electrode, thus leading to a non-zero tunnelling current. The coupling strength between the initial and final states in the tunnelling process is set by $|M^{s,s'}_{n,m, \xi}(\varepsilon)|^2$ which, because of the spinorial nature of the MLG and BLG wave functions, is expressed as a sum of four terms, each of which contains an integral $\int \tilde{\phi}_n^* \phi_m dA$ of two oscillator states $\tilde{\phi}_n$ and $\phi_m$. For ideal alignment of the electrodes, $\theta = 0^{\circ}$, the set $\{\tilde{\phi}_n\}$ is equivalent to $\{\phi_m\}$ and the integrals express orthonormality of functions with different indices, $\int \phi_n^* \phi_m dA = \delta_{n,m}$. As a result, tunnelling only occurs if one of the four conditions is fulfilled: i) $n=m$, ii) $n-1=m$, iii) $n=m-2$ and iv) $n-1=m-2$.  
Hence, the central region of large current in Fig. 2(a), for $V_g = 0$ and non-zero $V_b$ [finger-like features across the regions $(\mu_\text{MLG},\mu_\text{BLG})=(2,-4)$ and $(-2,4)$], corresponds to the coupling between $m=0$ and $n=0$ Landau levels in BLG and MLG respectively.  Although at low voltages the $m=0$ and $m=1$ Landau levels in BLG are degenerate,  transitions between the $m=1$ and $n=0$ level in MLG are forbidden. Moreover, although increasing $V_b$ increases the size of the tunnelling energy window to include higher Landau levels, due to the selection rules for $\theta=0$, these do not contribute to the tunnelling current.

Setting non-zero $V_g$, at constant $V_b$ dopes the graphene electrodes, shifting the two chemical potentials together such that the difference between them remains unchanged.  At small $V_b$ and zero $V_g$ in Fig. 2(a) clear current is observed. However, as $V_g$ is increased (decreased), the electrodes become hole-doped (electron-doped) and the filling factors are changed to (-2,-4) [(2,4)]. As a result, the tunnelling energy window, set by $V_b$, moves away from the positions of the $m=0$ and $n=0$ Landau levels and the current decreases. Additionally, $V_b$ and $V_g$  induce an electric field between the graphene layers in BLG which leads to non-zero interlayer asymmetry $u$. This opens a band gap in the electronic spectrum of BLG \cite{mccann_prl_2006} and hence affects the current characteristics of the device. Within the voltage window shown in panel (a), $u$ is the largest in the top right/bottom left corners of the $(V_b,V_g)$ diagram and reaches the magnitude of $\sim20$ meV.
 
Finally, the current diagram as shown in panel (a) seems to have inversion antisymmetry with respect to the point $(V_b,V_g)=(0,0)$, $I(V_b,V_g) = - I(-V_b,-V_g)$. In fact, within the voltage window presented in Fig. 2, this antisymmetry is only weakly broken by the energy dependence of the decay coefficient $c(\varepsilon)$ (see also Appendix A) - a feature also observed experimentally \cite{britnell_nl_2012, britnell_sci_2012}. We investigate this symmetry in more detail in panel (e) where we present current plots for changing $V_b$ and constant $V_g$ from -9 V to 9 V in steps of 3 V corresponding to solid/dashed lines marked in (a). We show with solid lines current for negative $V_g$ and $V_b$ and, with dashed lines, current for positive $V_b$ and $V_g$. The same colour is used for curves with the same magnitude of $V_g$ and, for all $V_g$, we have $I(V_g,V_b)$ almost equal to $-I(-V_g,-V_b)$.

In Fig. 2(b), we show the tunnelling current as a function of $V_b$ and $V_g$ for $V_t = 0.5$ V. Non-zero $V_t$ induces interlayer asymmetry, $u$, even for $V_b=V_g=0$ V while also introducing a shift between the MLG and BLG neutrality points, $\Delta$. The former leads to valley splitting resulting in new filling factor regions with $\nu_\text{BLG} = 0$, while the latter leads to energy misalignment of the $m=n=0$ Landau levels. However, because in BLG the position of the $m=0$ Landau level depends linearly on $u$ [see Eq. (7)], the impact of $V_t$ can be counterbalanced by choosing $V_g$ such that the overall $u$ shifts the $m=0$ BLG Landau level in the $K_+$ valley back into alignment with the $n=0$ Landau level in MLG. This restores the finger-like feature, in panel (b) visible for some positive $V_g$. 

In Fig. 2(c) and (d), we show the impact of misalignment $\theta=0.25^{\circ}$ and $\theta=0.5^{\circ}$, respectively, between the two electrodes on the tunnelling current.  For non-zero $\theta$, the oscillator functions $\tilde{\phi}_n$ and $\phi_m$  are no longer orthonormal and transitions between any pair of states are allowed. Moreover, as shown in Appendix B, the coupling strength between states also depends on the misalignment angle. For these reasons, in panels (c) and (d) the finger-like feature present in (a) and (b) becomes increasingly smeared out with increasing $\theta$ and the tunnelling current also decreases as compared to (a) and (b). Additionally, misalignment between the electrodes breaks the approximate inversion antisymmetry of the current diagram in (a). In the presence of $\theta\neq 0^\circ$, each of the four terms of the kind $\int\tilde{\phi}_{n}\phi_m dA $ appearing in the calculation of the matrix element $M_{n,m,\xi}^{s, s'}(\varepsilon)$ (see also Appendix B) comes with a prefactor that depends on the MLG and BLG band indices $s'$ and $s$.  Upon inversion from $I(V_g, V_b)$ to $I(-V_g, -V_b)$, the interference between these terms leads to different results depending on whether the initial and final states originate in the conduction or valence band. As a consequence, the approximate inversion antisymmetry about $(V_g,V_b)=(0,0)$, present in Fig. 2(a) is strongly broken in both (c) and (d). This is demonstrated in more detail in Fig. 2(f), where we show similar current curves as in panel (e) (changing $V_b$ for constant $V_g$ from -9 V to 9 V in steps of 3 V with the same colour scheme) produced for $\theta=0.25^{\circ}$ (the cuts are also indicated in panel (c)). In particular, the magnitude of the current for $V_g=0$ V is much larger for $V_b<0$ V than $V_b>0$ V.

    \subsection{Valley Polarisation at $B=1$ T}
     \begin{figure} [!h]
\subfloat{  \includegraphics[clip,width=\columnwidth]{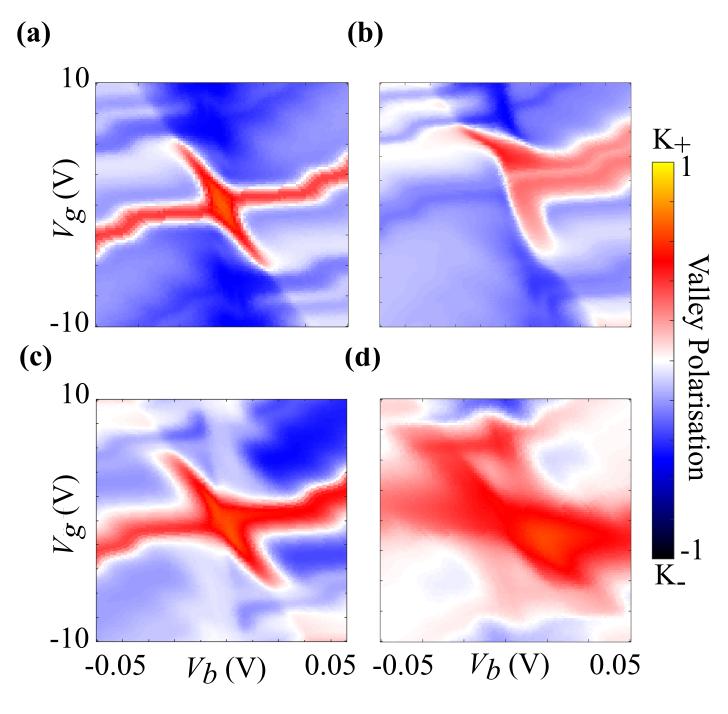} }
\caption{(Color online) Valley polarisation of the tunnellling current at $B = 1$ T, as a function of the bias and gate voltages $V_b$ and $V_g$. Each of the panels (a)-(d) corresponds to the total current shown in the equivalent panels (a)-(d) of Fig. \ref{fig:figure2}: (a) $\theta=0^\circ$  and $V_t = 0$ V (b) $\theta=0^\circ$ and $V_t = 0.5$ V, (c) $\theta=0.25^\circ$  and $V_t = 0$ V and (d) $\theta=0.5^\circ$  and $V_t = 0$ V. Positive (negative) polarisation indicates current favouring the $\vect{K_+}$ ($\vect{K_-}$) valley. }
\label{fig:figure3}
\end{figure}
 As the Landau level wave functions of BLG are not distributed equally between its two constituent graphene layers and this distribution is reversed between the valleys, tunnelling in the device shown in Fig. 1 can be used to produce unequal electron occupations in the MLG drain electrode. Such an effect can be characterised by the valley polarisation, $P$, of the tunnelling current, introduced in Eq. (13), which we plot in Fig. 3 as a function of the gate voltages $V_g$ and $V_b$ for $B=1$ T. Panels (a)-(d) in this figure correspond to the same parameters for which we presented total tunnelling current in Fig. 2(a)-(d).
 
 In panel (a), we show the case of $\theta=0^{\circ}$ and $V_t=0$ V. The polarisation diagram has inversion symmetry with respect to $(V_b,V_g)=(0,0)$. A bright red cross-like feature corresponds to $P \sim50\%$ with a region of $P\sim80\%$ in the centre of the diagram. This high valley polarisation is due to the tunnelling between the $m=0$ BLG and $n=0$ MLG Landau levels. In the BLG $\vect{K_+}$ valley, electrons in the $m=0$ state occupy exclusively the layer closer to the barrier, whereas in the $\vect{K_-}$ valley all of them sit on the layer further from the barrier. Consequently, the current in the $\vect{K_+}$ valley is significantly larger than in the $\vect{K_-}$ valley.
 
 As $V_g$ is increased, the $m=0$ and $n=0$ Landau levels move out of alignment and the dominant source of tunnelling current becomes the $m=2$ to $n=0$ transition. From Eq. (8), the BLG $m=2$ wave function is $\psi_{2,s}^{\xi}\propto(C_1 \phi_2, C_2 \phi_0)^T$, where $C_1$ and $C_2$ are complex numbers, and the first and second components of  $\psi_{2,s}^{\xi}$ are located respectively on layer 1 (layer 2) and layer 2 (layer 1) in valley $\vect{K_{+}}$ ($\vect{K_{-}}$). For the MLG $n=0$, $\tilde{\psi}=(\tilde{\phi_{0}},0)$, so that for $\theta=0^\circ$, tunnelling in $\vect{K_{+}}$ is only possible for BLG electrons from layer 2, further from the barrier, while in $\vect{K_{-}}$ it is the electrons from layer 1 that can tunnel into MLG. As a result, the overall current has negative ($\vect{K_-}$) polarization as  shown by dark blue regions above and below the central red cross in Fig. 3(a). Similar arguments can be used to explain other regions of the polarisation diagram.


For non-zero top gate voltage, as shown in Fig. \ref{fig:figure3}(b) for $V_t=0.5$ V, the polarisation map is modified as a result of the shift between the neutrality points, $\Delta$, as well as non-zero interlayer asymmetry, $u$, at ($V_b,V_g$)=(0,0). The latter lifts the valley degeneracy of the BLG $m=0$ Landau level. Alignment of the $\vect{K_+}$ BLG $m=0$ and MLG $n=0$ states, responsible for the red cross-like feature in (a), now requires compensating with positive gate voltage. However, for negative $V_b$ it is not possible to both align these two states and position the BLG and MLG chemical potentials such that the aligned states contribute to the current. As a result, the left arm of the red cross disappears and the $m=2$ to $n=0$ transitions lead to negative polarisation in this region. 

In Fig. \ref{fig:figure3}(c) and (d), we show valley polarisation  as a function of $V_b$ and $V_g$ for increasing misalignment between the electrodes, $\theta = 0.25^\circ$ and $\theta = 0.5^\circ$ corresponding to total current plots in Fig. 2(c) and (d). Similarly to the current features, when the graphene electrodes are misaligned,  individual polarisation features become smeared out and the variation of polarisation throughout the ($V_b,V_g$)-space becomes more gradual. The oscillator states $\phi_{m}$ and $\tilde{\phi}_{n}$ are not orthonormal for $\theta\neq 0^{\circ}$ so that many different transitions contribute to the overall polarisation for given ($V_b,V_g$). Importantly, interference of electronic states tunnelling between any of the BLG layers and any of the MLG sublattices which leads to different outcomes for conduction band-conduction band and valence band-valence band transitions, strongly breaks the inversion symmetry of polarisation present in Fig. 3(a) for $\theta=0^{\circ}$. This symmetry breaking grows with increasing $\theta$. 


	\subsection{Tunnelling at $B=4$ T}
The Landau level structures in BLG and MLG depend on the strength of the magnetic field differently, hence the tunnelling current and polarisation features in the ($V_b,V_g$) diagrams depend on $B$. For this reason, to contrast our results for $B=1$ T presented in Fig. 2 and 3 with the case of stronger magnetic field, in Fig. 4 we show the tunnelling current and its valley polarisation for $B=4$ T, $\theta=0^{\circ}$ and $V_t=0$ V. Due to the increased electron density per Landau level at $B=4$ T, it is necessary to increase the voltage range in order to compare features arising from similar electronic tunnelling transitions. 
Similarly to the $B=1$ T case, a finger-like structure is present across the regions $(\mu_\text{MLG},\mu_\text{BLG})=(2,-4)$ and $(-2,4)$ in Fig. 4(a). In fact, it is more pronounced because the separation between the $m=n=0$ Landau levels and the rest of the electronic spectra in the corresponding materials is increased. Consequently, the central cross-like region of $\vect{K_+}$-polarised current is also sharper, including polarisation of $P\sim90\%$ in the vicinity of $(V_b,V_g)=(0,0)$. The maximum $\vect{K_-}$ polarisation in the blue region dominated  by $m=2$ to $n=0$ tunnelling is also increased.
\begin{figure}
\subfloat{  \includegraphics[clip,width=\columnwidth]{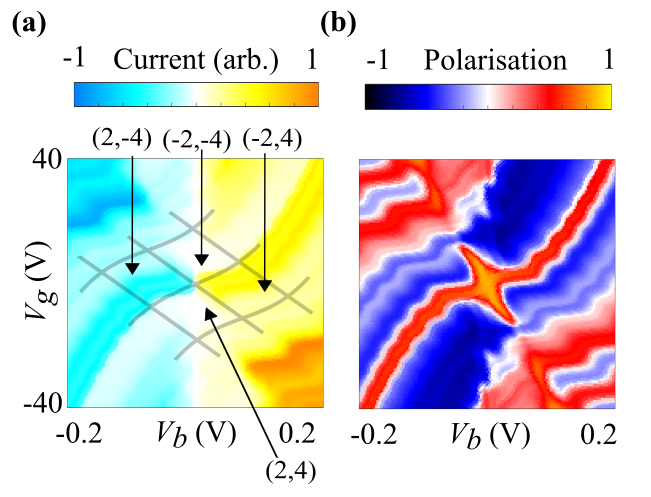} }
\caption{(Color online) Total tunnelling current (a) and valley polarisation (b) as a function of the bias and gate voltages $V_b$ and $V_g$ for magnetic field $B=4$ T, $\theta=0$ and $V_t=0$ V. The maximum polarisation observed is as much as 90$\%$ in favour of the $\vect{K_+}$ valley.  Grey lines [in (a)] indicate the boundaries of regions of constant filling factors labelled  $(\nu_\text{MLG}, \nu_\text{BLG})$, with the first (second) filling factor corresponding to monolayer (bilayer) graphene.  The filling factor regions in (b) are identical to those in (a).
}
\label{fig:figure4}
\end{figure}

\section{Tunnelling with strong momentum scattering}

In the presence of a poor interface between the electrodes and the hBN barrier, the scattering length-scale becomes very small such that the momentum resolution of the tunnelling electron becomes lost. In this limit, the momentum-nature of the initial and final state has no effect on the magnitude and valley polarisation of the current across the device. Instead, the tunnelling current depends only on the density of states of the source and drain electrode. As a consequence, we expect the valley polarisation of the tunnelling current to arise purely due to differences in valley occupations of the two BLG layers. We model this regime by setting each harmonic oscillator integral, $\int \tilde{\phi}_n \phi_m dA$, equal to 1 for all the transitions independently of their initial and final states \cite{Note2} and present our results for $\theta=0^{\circ}$ and $V_t=0$ V in Fig. 5.

\begin{figure} [!h]
\subfloat{  \includegraphics[clip,width=\columnwidth]{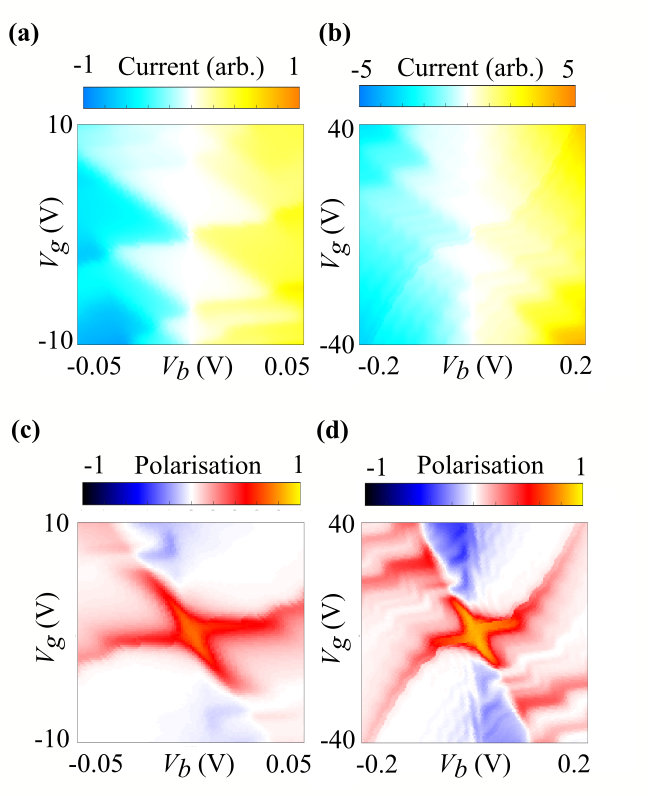} }
\caption{(Color online)  Electron transport through the proposed device in the absence of momentum conservation and as a function of the bias and gate voltages $V_b$ and $V_g$ with $ V_t=0$. Panels (a) and (b) show the total tunnelling current for $\theta$=0$^{\circ}$ and (a) $B=1$ T, (b) $B = 4$ T. Panels (c) and (d) present valley polarisation of the current shown in (a) and (b), respectively.    }
\label{fig:figure5}
\end{figure}

In panels (a) and (b), we show the tunnelling current for $B=1$ T and $B=4$ T, respectively. Because all of the transitions are now allowed, the graphs look similar to that in Fig. 2(d), corresponding to $\theta=0.5^{\circ}$. In (b), the increased magnitude of the magnetic field leads to larger spacing between the Landau levels so that, as compared to (a), a larger voltage window is necessary to capture features due to transitions between the same pair of Landau levels.

In Fig. 5(c) and (d), we present valley polarisation of the currents shown in Fig. 5(a) and (b), respectively. The relaxation of the selection rules discussed in the previous section results in polarisation maps which are heavily weighted in favour $K_+$ valley. In particular, the maximum valley polarisation occurs at low voltages, where the participating Landau levels are those with low index (in particular $m=0$ and $m=1$). This is because, for the $m=0$ and $m=1$ BLG Landau levels, the valley and layer degrees of freedom are coupled. Furthermore, the interlayer asymmetry, $u$, generates a layer population difference in the $m \geq 2$ Landau levels in BLG, which is opposite in the two valleys. This induced interlayer asymmetry is responsible for small regions of minor $K_-$ polarisation  which occur at higher voltages. These two principles are responsible for all polarisation features observed in Fig. 5(c) and (d).

The relative misalignment of the graphene electrodes has no effect on the tunnelling probability in this limit.   Similarly to the case of momentum-conserving tunnelling between misaligned electrodes, $\theta\neq 0^{\circ}$, lack of restrictions on allowed transitions leads to chiral interference. This interference results in the asymmetry in inversion about the origin that is observed in Fig. \ref{fig:figure5}.
We expect momentum non-conserving tunnelling to be the dominant mechanism when the misalignment angle between the graphene electrodes is large. This is because, while increasing misalignment angle decreases the magnitude of the  momentum-conserving current, it should have no effect on the transitions involving scattering.


\section{Summary}

We have explored the tunnelling characteristics of a vertical field effect transistor comprising monolayer and bilayer graphene electrodes, in the presence of a perpendicular magnetic field.  The coupled layer and valley polarisation in the Landau levels of bilayer graphene gives rise to a valley-polarised tunnelling current through the device resulting in unequal valley populations in monolayer graphene. Our result is due to the difference in effective tunnelling barrier widths for electrons in the two layers of the BLG electrode. As such, valley polarisation should persist in the presence of small local variations of the tunnelling rates (and hence effective tunnelling decay lengths). Importantly, this valley polarisation can be tuned solely by electrostatic means without the need to reverse the direction of the magnetic field. Our modelling suggests that $P\sim\pm 70\%$ is possible in high quality devices in homogeneous fields of $B=1$ T. Fields of such magnitude could be in principle generated by placing ferromagnets on top of the device \cite{Nogaret}. While the homogeneity of the field distribution across the device would then depend on the size of the ferromagnet, thickness of the tunnelling junction and distance between the two, valley polarisation might still be possible in such a setup.  

In both the momentum-conserving and non-conserving regimes, the most persistent feature in valley polarisation plots is the cross-like region of $\vect{K_+}$-polarised current around $(V_g, V_b) = (0,0)$. In the same voltage region, the total tunnelling current forms a finger-like pattern. Both originate in tunnelling current from $m=0$ ($m=0,1$ in the absence of momentum conservation) to $n=0$, so that observing the finger-like features in the current should indicate a region of considerable valley polarisation.
In order to detect the valley polarisation produced using the proposed device directly, two stacks could be connected in series: the first one to produce unequal valley populations and the second to act as a detector. Alternatively, the produced valley polarisation can be measured using optical means \cite{Hipolito}.

\begin{acknowledgements}

The authors acknowledge helpful discussions with E. McCann and K. Takashina. This work has been supported by the EPSRC through the University of Bath Doctoral Training Partnership, grant numbers EP/M50645X/1 and EP/M507982/1.
\end{acknowledgements}

\begin{appendix}
\section{Changing the decay constant of graphene}
In the main text, we take the decay constant $c'(\varepsilon)$ characterising tunnelling through monolayer graphene to be equivalent to the decay constant $c(\varepsilon)$ corresponding to tunnelling through hBN. Here we discuss the effect of changing the decay constant $c'(\varepsilon)$, on our results. 
Following previous work \cite{britnell_sci_2012, simmons_1963}, we relate the decay constant to the height of the tunnelling barrier. For the case of hBN, we treat it as an isotropic potential step  with barrier height $\Phi_0 = -1.5$ eV, corresponding to experimental measurements of the valence band maximum (VBM) of hBN  \cite{britnell_sci_2012, georgiou_natnano_2013}. The hBN energy dispersion around the VBM is roughly parabolic in $k_z$ and this allows us to write \cite{simmons_1963}

\begin{align} \label{decaycons}
    c(\varepsilon) = \text{Im} \dfrac{\sqrt{2 m^* \Phi(\varepsilon)}}{\hbar} =  \text{Im} \dfrac{\sqrt{2 m^* (\Phi_0-\varepsilon)}}{\hbar} , 
\end{align}
where $m^*$ is the effective mass. The above relation predicts weak electron-hole asymmetry in tunnelling current (as observed in experiments \cite{britnell_sci_2012, georgiou_natnano_2013}). In our work, we use the expression in Eq. (A1) to obtain the decay constant $c'(\varepsilon) = c(\varepsilon) $ for the tunnelling of BLG electrons from the layer further from the barrier across the graphene layer closer to the barrier. While both the barrier height and the effective mass would be different for graphene as compared to hBN (here, for the sake of the numerical calculations, for hBN we take $m^*=0.5m_0$ following previous modelling of vertical tunnelling in graphene/hBN stacks  \cite{britnell_sci_2012, georgiou_natnano_2013}), our main conclusions are quite insensitive to the numerical values of $c'(\varepsilon)$ and $c(\varepsilon)$. In fact, the latter impacts both the electrons tunnelling from the top and bottom BLG layers in the same way and hence leads to an identical numerical coefficient for all tunnelling processes for given applied voltages. The physics we describe arises primarily due to the additional exponential factor, $\exp{(-c'(\varepsilon)d_0)}$, in tunnelling from the bottom layer as compared to the top one.  

\begin{figure}
\subfloat{  \includegraphics[clip,width=\columnwidth]{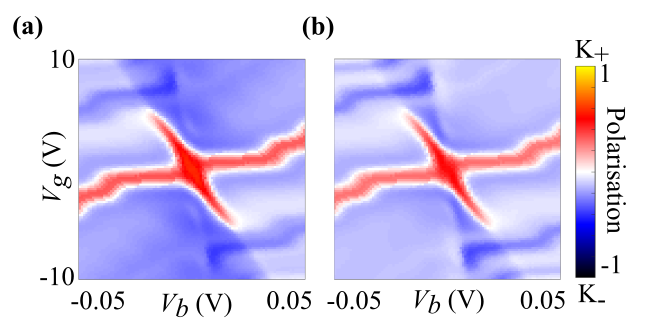} }

\caption{(Color online) Valley polarisation for momentum-conserving tunnelling between perfectly-aligned electrodes ($\theta=0^\circ$) and for $B=1$ T as a function of $V_b$ and $V_g$ with $V_t = 0$ V. In contrast to the polarisation shown in Fig. 3(a), the plots here are obtained using the decay constant decreased by a factor (a) $\frac{1}{\sqrt{2}}$ and (b) $\frac{1}{2}$ (compared to in the main text). }
\label{fig:figure7}
\end{figure}

In  Fig. (\ref{fig:figure7}), we demonstrate the valley polarisation at $B=1$ T and $\theta = 0^\circ$ for effective graphene decay constant, $c'(\varepsilon)$, scaled by a factor of (a) $\frac{1}{\sqrt{2}}$ and (b) $\frac{1}{2}$ as compared to the hBN value provided by Eq. (A1). While the asymmetry between $P(V_b,V_g)$ and $P(-V_b,-V_g)$ increases slightly for smaller decay constant, qualitative features of the valley polarisation graphs remain the same. Also, the maximum valley polarisations are still significant, $58\%$ and $48\%$ respectively, compared to $~70\%$ in Fig. 3(a). Increasing $c'(\varepsilon)$ (making graphene more insulating) increases the valley polarisation of the tunnelling current.

\section{Landau Level Couplings and Matrix Element in the Momentum-Conserving Limit}
\captionsetup[subfigure]{position=top, labelfont=bf,textfont=normalfont,singlelinecheck=off,justification=raggedright}
 \begin{figure} [!t]
\subfloat{  \includegraphics[clip,width=\columnwidth]{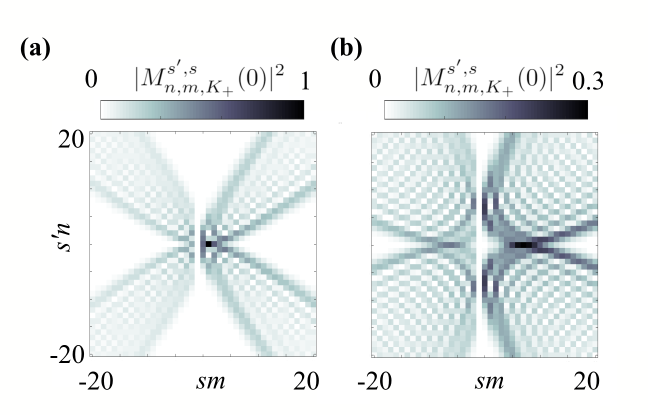} } 
\caption{(Color online) Colour map of the tunnelling matrix element $|M_{n,m, K_+}^{s,s'}(0)|^2$ between Landau levels of MLG (with indices $s'n$) and gapless BLG (indices $sm$) for $B=1$ T and (a) $\theta=0.25^{\circ}$ and (b) $\theta=0.5^{\circ}$. All values are normalised to the maximum value in (a). }
\label{fig:figure6}
\end{figure}
The matrix element determining the tunnelling between the BLG and MLG electrodes depends on the Landau level indices, $n,m$, as well as the  magnetic field, $B$, and misalignment angle between the two sheets, $\theta$. For the $\vect{K_+}$ valley, it can be written as  
\begin{align} 
 &M_{n,m,K_{+}}^{s',s}(\varepsilon) =  \frac{V_{0}P_{n,m}}{\sqrt{2}} e^{-c(\varepsilon) d } \\ \notag 
& \times \bigg[ C_{1}\Lambda^\xi_{n,m} - s' i e^{i \theta} C_{1} \Lambda^\xi_{n-1,m} \\\notag
 & + e^{-c'(\varepsilon) d_0 } \left(  C_{2} \Lambda^\xi_{n,m-2}- s' ie^{i \theta} C_{2} \Lambda^\xi_{n-1,m-2} \right) \bigg],
 \end{align}
 whereas, for the $\vect{K_-}$ valley, we obtain
 \begin{align}
 &M_{n,m,K_{-}}^{s',s}(\varepsilon) = \frac{V_{0}P_{n,m}}{\sqrt{2}}e^{-c(\varepsilon) d } \\ \notag
 & \times \bigg[ C_{2} \Lambda^\xi_{n,|m|-2} 
+s' i e^{-i\theta} C_{2} \Lambda^\xi_{n-1,m-2} \\ \notag
& + e^{-c'(\varepsilon) d_0 } \left( C_{1} \Lambda^\xi_{n,m} + s' i e^{-i\theta} C_{1} \Lambda^\xi_{n-1,m} \right) \bigg]. \end{align}
In both cases we define
\begin{align} \notag
 &C_{1} =\begin{cases} \dfrac{\varepsilon^0_{m}}{\sqrt{C}}  &  \quad \quad \quad \quad \quad \quad \quad  \quad \, \ \ m\neq 0,1 \\
 1  &  \quad \quad \quad \quad \quad \quad \quad  \quad \, \ \  m =  0,1
 \end{cases},
 \\ \notag
&C_{2} = \begin{cases}\dfrac{\left[\varepsilon_{m,s,\xi}- \xi \frac{u}{2} +\xi u m \eta^2\right]}{\sqrt{C}} & m\neq 0,1\\ 0 & m = 0,1 
 \end{cases}, \\  \notag
&\Lambda^\xi_{n,m} = N_{n,m}2^{\text{max}\{n, m\}} (\text{min}\{n,m\})! e^{i\frac{1}{2} \Delta K_{\xi}^{x} \Delta K_{\xi}^{y} \lambda_B^2} \\ 
\notag & \times                                                                    \left(\text{sgn}(n-m) \frac{1}{2} \lambda_{B}\Delta K_{\xi}^{y} - i \frac{1}{2} \lambda_{B} \Delta K_{\xi}^{x})\right)^{|n-m|} \\
&\times e^{- \dfrac{\vect{\Delta K_\xi}^{2}\lambda_{B}^{2}}{4}} \mathcal{L}_{\text{min}\{n,m\}}^{|n-m|}\left( \dfrac{\vect{\Delta K_\xi}^{2}\lambda_{B}^{2}}{2} \right),
\end{align}
where $N_{n,m}$ and $P_{n,m}= \sqrt{(1+\delta_{n,0})}$ are normalisation constants and $\mathcal{L}_{\alpha}^\beta (x)$ are generalised Laguerre polynomials.
The strength of the coupling  at the $\vect{K_+}$ valley, $| M_{n,m,K_{+}}^{s',s}|^2$, is shown for $B=1$ T and as a function of Landau level, $n,m$, and band indices, $s,s'$, in Fig. \ref{fig:figure6}(a) and (b) at misalignment angles $\theta=0.25^\circ$ and  $\theta=0.5^\circ$ respectively.

For zero misalignment angle, the matrix element is simply a linear combination of Kronecker deltas (each expressing orthonormality of the harmonic oscillator states), suggesting only transitions between certain Landau states are allowed.  However, as shown in Fig. 7(a) , increasing  misalignment redistributes the coupling strength amongst other transitions (in particular, for any non-zero $\theta$ transitions between any two oscillator states are in principle allowed). Interestingly, changing the misalignment angle also changes the preferred transition (the one with the largest coupling strength). However, because the matrix element depends on products of the type $\vect{\Delta K_\xi}^2 \lambda_B^2$, a change of angle (which determines $\vect{\Delta K_\xi}$) can be to some extent counterbalanced by changing the magnetic field (and hence $\lambda_B$).

\section{Electrostatics}
The bias voltage $V_b$ and gate voltages $V_g$ and $V_t$ control the  local Fermi levels in BLG and MLG as well as the shift between the neutrality points and the interlayer asymmetry in BLG.  We use a four-plate capacitor model to express the electric fields between the gates and consecutive graphene layers (we treat hBN and SiO$_2$ as homogeneous insulators with dielectric constants $\epsilon_\text{hBN}$ and $\epsilon_{\text{SiO}_2}$, respectively). The carrier densities per graphene layer on the BLG source ($j=$BLG) or MLG drain ($j=$MLG) electrodes can be expressed as
\begin{align}
& n_{j} =\frac{1}{\pi^2 \lambda_B^2} \nonumber \\ & \times \! \! \sum_{m,s,\xi}\left[\arctan{\left(\frac{\mu_{j}-\varepsilon^\xi_{m,s}}{\Gamma_{j}}\right)} -\arctan{\left(\frac{-\varepsilon^\xi_{m,s}}{\Gamma_{j}}\right)}\right].
\end{align}
Through charge conservation, for each combination of $\mu_\text{BLG}$, $\mu_\text{MLG}$ and $u$, we obtain corresponding bias, bottom  and top gate potentials. Furthermore, the charge build up on the bilayer sheet acts as a capacitance leading to a difference in neutrality points of the two spectra
\begin{equation}
\Delta=  E^{\text{MLG}}_0-E^{\text{BLG}}_0. 
\end{equation}
For simplicity we set $E^{\text{BLG}}_0 = 0$ and therefore the position of the charge neutrality point in monolayer is related to the number of excess charge carriers in the bilayer graphene by relation
\begin{align}
E^{\text{MLG}}_0 =\Delta = -\frac{e^2 (n_{\text{Au}} + n_{\text{BLG}}) d}{\epsilon_0 \epsilon_{\text{hBN}}}. 
\end{align}
The voltages of the system can therefore be expressed as
\begin{align}
&V_{b} = \frac{1}{e} \left[ \mu_\text{BLG} - \mu_\text{MLG} - \Delta \right], \nonumber \\ 
 &V_g = -\frac{e(n_\text{MLG}\!+\!n_\text{BLG}+\!n_{\text{Au}})(d_{\text{SiO}_2}\epsilon_{\text{hBN}}\!+\!d_{\text{hBN}}\epsilon_{\text{SiO}_2})}{\epsilon_{\text{hBN}}\epsilon_{\text{SiO}_2}\epsilon_0}, \nonumber\\
   & V_t = \frac{-en_\text{Au}d_{\text{Top}}}{\epsilon_0 \epsilon_\text{hBN}},
\end{align}

\end{appendix}


\begin{thebibliography}{99}
\bibitem{note1}
The single-sublattice occupation of the $n=1$ BLG Landau level is not preserved beyond the low-energy two-band model we use here \cite{papermarcin}. However, it remains a good approximation for the magnetic fields considered here.
\bibitem{footnotell}
Because at $B=1$ T, the Landau level separation in BLG, $\eta^2 \gamma_1 \approx 3.5$meV, is comparable to our Landau level broadening of 2 meV, we draw the boundaries between regions by simply rounding the electron density to the nearest filling factor.
\end{thebibliography}
\end{document}